# Observation of Three-dimensional Photonic Dirac points and Spin-polarized Surface Arcs

Qinghua Guo[1, 2†], Oubo You[1, 3†], Biao Yang[1, 4†], James B. Sellman[1], Edward Blythe[1], Hongchao Liu[1], Yuanjiang Xiang[3], Jensen Li[1, 2], Dianyuan Fan[3], Jing Chen[5*], C. T. Chan[2*], Shuang Zhang[1*]

1. School of Physics & Astronomy, University of Birmingham, Birmingham, B15 2TT, UK

2. Department of Physics and Center for Metamaterials Research, The Hong Kong University of Science and Technology, Hong Kong, China.

3. International Collaborative Laboratory of 2D Materials for Optoelectronic Science and Technology of Ministry of Education, Shenzhen University, Shenzhen 518060, China.

4. College of Advanced Interdisciplinary Studies, National University of Defense Technology, Changsha 410073, China.

5. School of Physics, Nankai University, Tianjin 300071, China.

*Correspondence to: jchen4@nankai.edu.cn; phchan@ust.hk; s.zhang@bham.ac.uk

†These authors contributed equally to this work.

**Topological phases arise from the elegant mathematical structures imposed by the interplay between symmetry and topology[1-5]. From gapped topological insulators[1,2,6-9] to gapless semimetals [4,10-25], topological materials in both quantum and classical systems, have grown rapidly in the last decade. Among**

them, three-dimensional Dirac semimetal[26-29] lies at the topological phase transition point between various topological phases. It shares multiple exotic topological features with other topological materials, such as Fermi arcs and chiral anomaly with Weyl semimetals[30], spin-dependent surface states with topological insulators[29]. In spite of the important role it plays in topological physics, no experimental observation of three-dimension Dirac points has been reported in classical systems so far. Here, we experimentally demonstrate three-dimension photonic Dirac points in an elaborately designed photonic metamaterial, in which two symmetrically placed Dirac points are stabilized by electromagnetic duality symmetry[31]. Spin-polarized surface arcs (counterparts of Fermi arcs in electronic systems) are demonstrated, which paves the way towards spin-multiplexed topological surface wave propagation. Closely linked to other exotic states through topological phase transitions[32,33], our system offers an effective medium platform for topological photonics.

Three-dimensional (3D) Dirac points inheriting relativistic effects from high-energy physics appear as gapless excitations in the topological band theory. Their presence usually requires extra constraints, such as spatial symmetries [8,9,26,28,34-36], where the Dirac state space is constructed by spatial Bloch modes. In photonics, when the wavelength of electromagnetic waves approaches to long wavelength limit, effective medium theory can be employed to describe the optical properties of the structured media[37-39]. It has been shown that the intrinsic polarization states can be used to construct the Dirac state space in judiciously designed photonic metamaterials for realizing three-dimensional Dirac degeneracies[40].

A 3D Dirac point is composed of two overlapping Weyl points with opposite topological charges. In metamaterials, Weyl points can be realized by the crossing between a longitudinal mode and a circularly polarized transverse mode, where the Weyl point charge is given by the spin of the circular transverse mode [22,23,41]. Here, we construct stable 3D photonic Dirac points by further imposing electromagnetic duality symmetry[40]. It is an internal symmetry of electromangtic field[31]. In artificial metamaterials, duality symmetry requires the proportionality between permittivity and permeability tensors, such as $\overleftrightarrow{\varepsilon} = \eta \overleftrightarrow{\mu}$ ($\eta > 0$). Now the Dirac point state space is doubled by the electromagnetic duality symmetry. Compared with a single Weyl point, it is spanned by two longitudinal modes and two circular polarized transverse modes[40]. At the interface between air and the Dirac metamaterials, we observed the spin dependent topological surface states in photonics.

In order to simultaneously realize the electric and magnetic longitudinal modes, we utilize metallic helical elements to introduce both electric and magnetic resonances in the z direction, as shown in Fig. 1a. Meanwhile, in order to eliminate the unwanted overall chirality of the metamaterial, each unit cell consists of two layers of helical elements with opposite chirality: the four identical helical elements[42] on layer 1 (L1) rotating clockwise, while those on layer 2 (L2) rotating anti-clockwise. They are related to each other through the mirror operation as indicated by plane M. Further finely tuning the structure parameters of the helical elements leads to double degeneracy of the two longitudinal modes. Finally, the helical elements are arranged to preserve $C_4$ rotation symmetry in the x-y plane to ensure an in-plane isotropic response, and thus guaranteeing double degeneracy of the two circular polarized transverse modes propagating along z direction as well. In total, each unit cell consists of eight helical elements.

As shown by the simulated band structure (calculated using CST microwave studio) in Fig. 1d, two Dirac points (marked by the blue spheres) are symmetrically located on the $k_z$ axis at the same frequency around 7.92GHz. Due to the flat dispersion of the longitudinal modes, the tilted Dirac points lie right at the transition between type I and type II Dirac points, and they are therefore called transitional Dirac points. To clearly see the massless dispersion around Dirac points, Fig. 1e and f show the bands linearly threading through the Dirac points along $k_z$ and $k_x/k_y$, respectively. Along the

$k_z$ direction, the two longitudinal modes (blue solid and dashed lines) and the two transverse modes (black solid and dashed lines) are degenerate across the entire frequency range. However, the degeneracy is slightly lifted along $k_y$ direction away from the Dirac point, because the condition for ideal electromagnetic duality, i.e. exactly tensor-matching between permittivity and permeability, cannot exist in a wide frequency range (supplementary information 1, Fig. S1) in a metamaterial.

We fabricate the sample using Printed Circuit Board (PCB) technology. The clockwise/anticlockwise helical elements are printed on a 1mm-thick dielectric layer (with dielectric constant of 2.2) as shown in Fig. 1b. Between them, a spacer layer with thickness of 3mm and dielectric constant of 2.2 is placed to prevent short contacting between neighbor metallic elements. Finally, a 3D bulk metamaterial is constructed through layer-stacking (70 periods) along z direction as schematically shown in Fig. 1c, therein each period (8mm) exhibits L1-Spacer-L2-Spacer configuration. For each structured layer, there are 20 and 70 unit cells with identical period $a = 6.8mm$ along x and y, respectively. The realistic sample is shown in Fig. S2 (supplementary information 2).

We probe bulk states of the metamaterials using a transmission method. One antenna is placed at the center of bottom surface of the sample as source, and another probe antenna raster-scans the transmitted field on the top surface. After subsequent Fourier transformations, we obtain the projection of the bulk states on the $k_y$-$k_z$ plane. The projected band spectra along $k_z$ and $k_y$ directions are shown in Fig. 2a and c, respectively. The 3D Dirac points are observed around frequency 7.92GHz. Fig. 2b and d show the corresponding simulated results for reference, where linear dispersion along different directions confirms the presence of the photonic Dirac points. Besides the bulk states, one of the surface states is clearly shown as the bright line connecting the Dirac points and light cone (blue line) as shown in Fig. 2a. The simulation (Fig. 2b) shows two surface states, however, the other is hard to discern from the measurement results. As expected, away from the Dirac point the projected bulk states are gapped, which agrees well with the simulation as shown in supplementary information 3, Fig. S3.

Surface state arcs, corresponding to Fermic arcs in electronic sytems, are important topological signatures of Dirac points. In order to clearly measure the surface state arcs, we use a slightly different experimental configuration as shown in Fig. 1c: the source antenna is moved to the edge of the top surface for better excitation of surface waves. The spatial field in Fig. 1c shows the propagating electromagnetic wave measured in this configuration. Fig. 3a-c show the surface state arcs at frequency 7.8GHz, 7.92GHz and 8.1GHz respectively. They correspond to the frequency below, at and above the Dirac points.

At the Dirac point, two Fermi arcs emerge from the Dirac point and tangentially terminate at the air-circle. Since a Dirac point consists of two Weyl points with opposite charges, each Fermi arc from the corresponding Weyl point should be spin-polarized, as distinguished by the magneta and cyan colors in Fig. 3d-f. Further effective media analysis[40] shows that the two Fermi arcs exhibit left/right circular polarization (L/RCP) with the plane of the polarization parallels to the Fermi arcs. In order to verify the spin-dependent nature of the Fermi arc, we further use circular polarized antennas to selectively excite each spin-polarized surface state arc. As shown in Fig. 3g and i, under LCP/RCP excitation, the surface wave propagates towards $-/+$ y direction. The asymmetry clearly demonstrates the spin-momentum locking property of surface state arcs (Fig. 3 h and j) in the Dirac system. The spin resolved intensity of the surface state arcs at different frequencies could be seen in Fig. S4 (supplementary information 4). The launching and propagation of robust spin-momentum locked surface wave demonstrated here may pave a new way in information processing, such as spin- multiplexed channels.

Below the Dirac frequency, the equi-frequency contours (EFC) of the bulk states (black solid lines in Fig. 3d) form two almost-overlapping hyperboloids (more experimental results are shown in Fig. S5). As shown in Fig. 3a and d, the surface states still exist at the gap between the metamateiral bulk state and light cone. However, there is a small gap between the surface state as numerically shown in Fig. 3d, where the impedance mismatch between metamaterial and air mixes LCP and RCP Fermi arcs. Above the Dirac points (Fig. 3c and f), the EFCs of the surface states show strong resemblance to the famous Dyakonov surface waves in anisotropic crystals[43].

As a gapless topological phase, a Dirac semimetal can be viewed as a parent structure which can generate various interesting topological materials, e.g. topological insulators, Weyl semimetals and nodal lines, through symmetry reduction. For example, by introducing a bi-anisotropic term[6,9] into the y-z plane, the Dirac point is gapped and the material transforms to a photonic topological insulator as shown in Fig. 4b, where there are gapless interface states (red dashed lines) across the bulk gap. Figure 4g and h show that Weyl points emerge when either inversion (P)[22,44,45] or time-reversal (T)[41,46] symmetry is broken. Each Weyl state space is spanned by one electric/magnetic longitudinal mode and one LCP/RCP. When the state space consists of two orthogonal linear transverse modes as shown in Fig. 4i, one can realize topological nodal lines. Furthermore, several triply degenerate points (TDPs) appear as transient phases as shown in Fig. 4c-f. Among them, the TDPs in Fig. 4c and d consisting of two spinless longitudinal modes and one spinful circular polarized transverse mode are chiral and represent nontrivial spin-1 gapless states[33]. In Fig. 4e, the state space of TDPs is constructed by one longitudinal mode and two degenerate circular polarized transverse modes, which are neutral behaving as triply degenerate Dirac points, which can further split into two Weyl points by breaking either inversion or time-reversal symmetry (Fig. 4g and h). Neutral TDPs can also be formed by two longitudinal modes and one linear polarized transverse mode (transverse electric/magnetic, TE/TM), as shown in Fig. 4f. Hypothetical constitutive parameters that give rise to those topological phases are presented explicitly in Table S1.

In summary, we have experimentally demonstrated 3D Dirac degeneracies in a photonic metamaterials operating at microwave frequencies. The presence of 3D Dirac points has been confirmed by demonstrating both the bulk states and the spin polarized surface state arcs. The demonstrated Dirac degeneracies provide a platform for studying the transition between various different photonic topological states. Moreover, our demonstrated surface state spin-orbital coupling shows great potential in manipulating topological surface photonics. The slightly gapped surface states may also lead to observation of surface Zitterbewegung oscillation[47].

# Figures

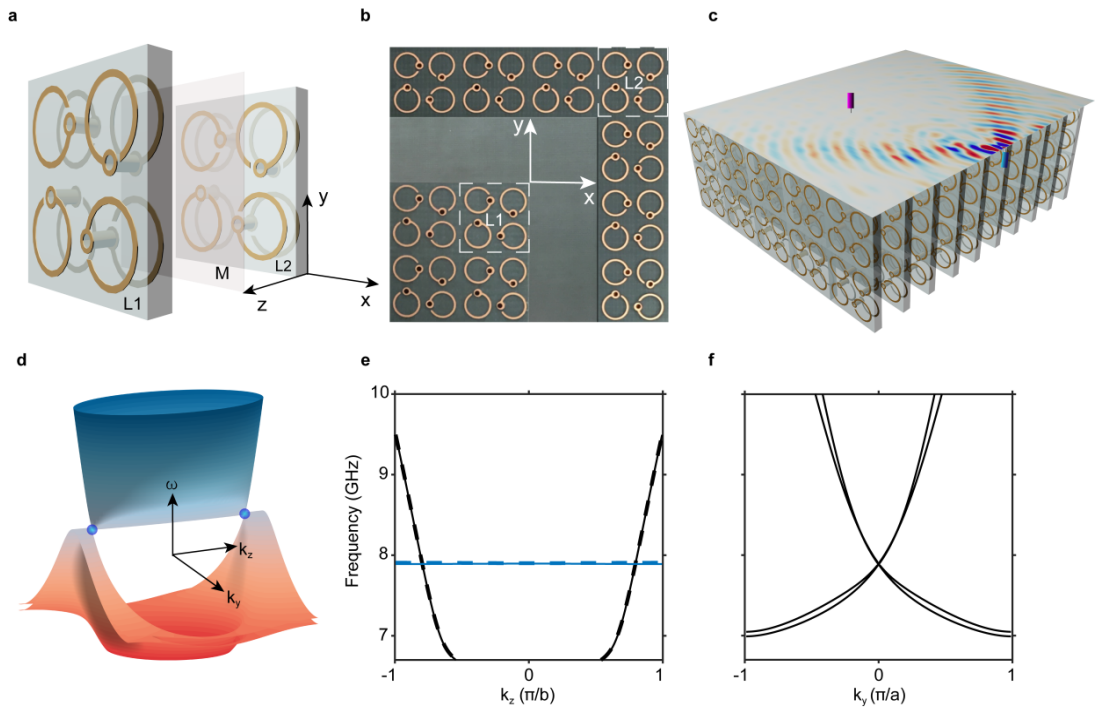

**Figure 1 | Three-dimension photonic Dirac metamaterial. a,** Schematic of the metamaterial structure. Each unit cell consists of eight helical elements. Along the z direction, there are two sets of helical elements related by mirror symmetry indicated by M (four clockwise and four anticlockwise rotated helical elements). In the x-y plane, the unit cell possesses $C_4$ rotation symmetry along z to ensure degeneracy between the two circular polarized transverse states. **b,** Sample fabricated with printed circuit board (PCB) technology. The helical elements are etched in the 1mm-thick hosting material (dielectric constant of 2.2). Between two structured layers there is a 3mm-thick spacer with the same dielectric constant. The in-plane period (marked as the white dashed squares) is $6.8 \times 6.8$ mm$^2$. The period along z is 8mm. **c,** Sample configuration for the experimental measurement. The sample is stacked layer by layer to form a three dimensional bulk metamaterial. There are two antennas, with one

working as the source (cyan) and the other working as probe (magenta). **d**, The presence of two four-fold degenerate Dirac points at the same frequency around 7.92GHz. **e,** The dispersion along $k_z$ directions around the Dirac points. **f,** Similar to **e** but along $k_y$.

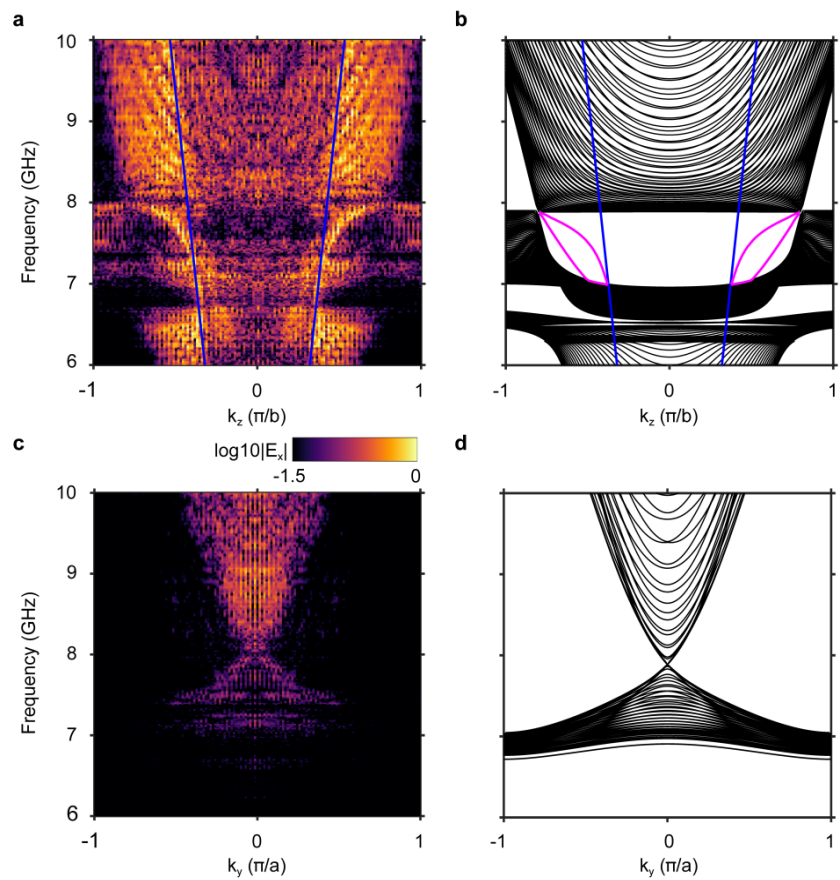

**Figure 2 | Observation of three-dimension photonic Dirac points**. **a** and **c,** Experimentally mapped band projection around the Dirac point along $k_z$ and $k_y$ direction, respectively. **b** and **d,** Corresponding simulation results from CST microwave studio. Blue and magenta lines indicate light cone and surface states, respectively.

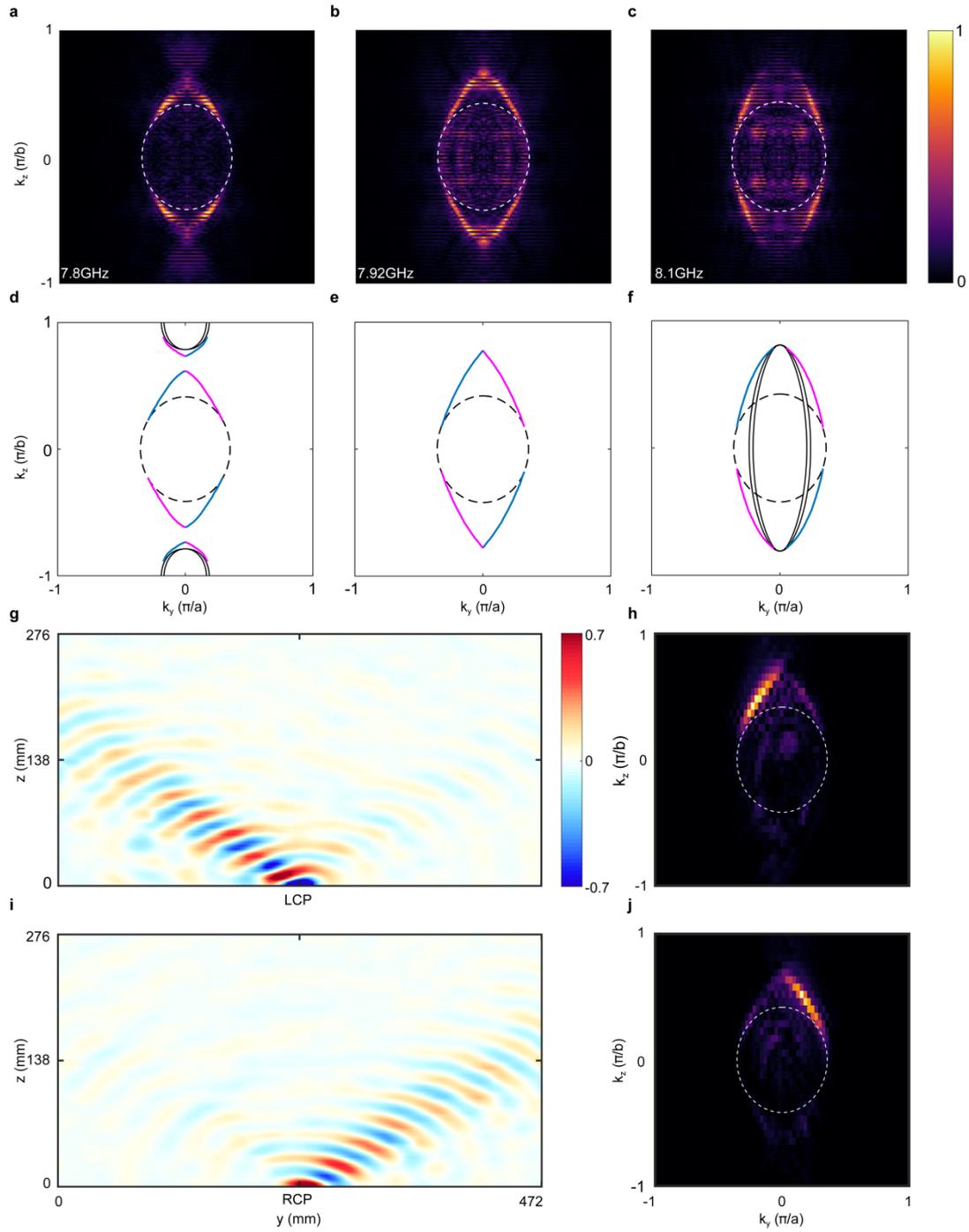

**Figure 3 | Spin polarized photonic surface arcs and surface waves. a-c** Experimentally mapped surface arcs at the frequency 7.8GHz (below Dirac point), 7.92GHz (at Dirac point) and 8.1GHz (above Dirac point), respectively. **d-f,** The corresponding simulated bulk and surface states, where cyan and magneta curves indicate opposite spinful surface states from CST microwave studio. Black solid lines

indicate simulated bulk states. **g,** Measured real space surface wave propagation excited by left circular polarization. **h,** Spin-polarized surface states in momentum space after Fourier transformation from **g**. **i-j,** Similar to **g-h** but excited by right circular polarization. In each panel, dashed circle indicates light cone.

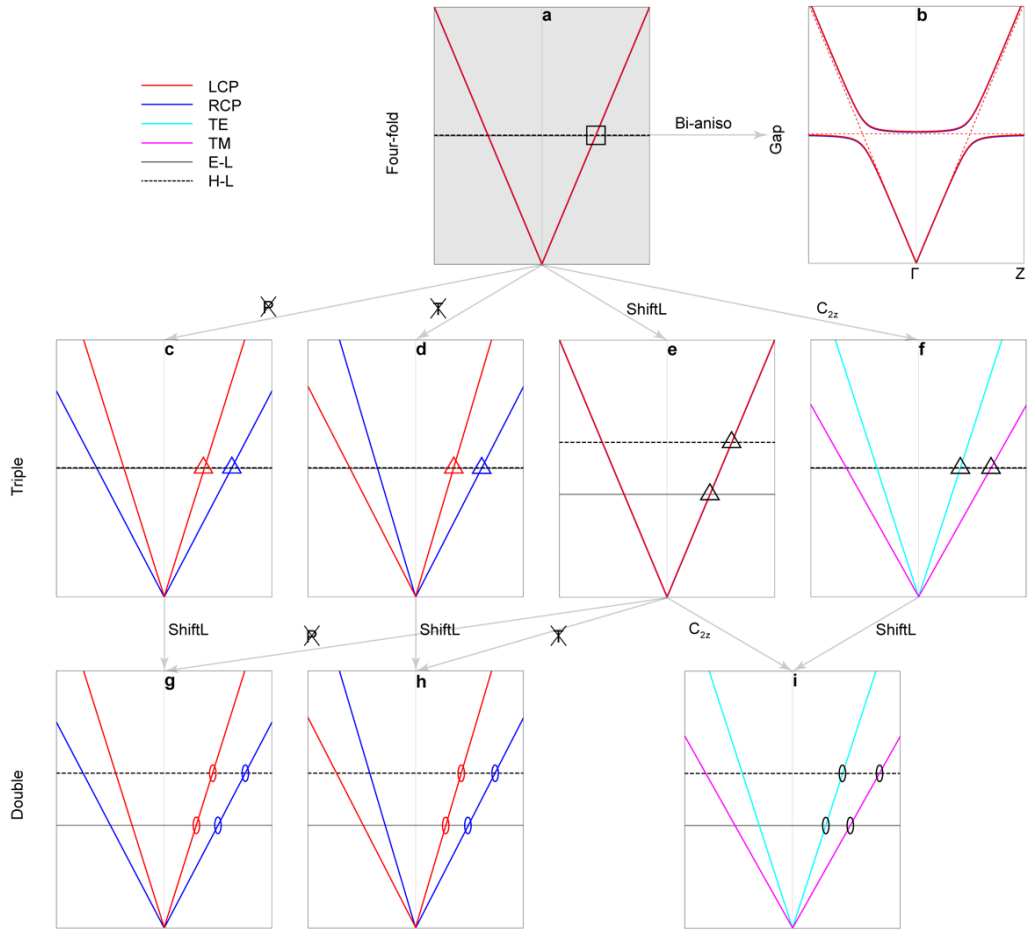

**Figure 4 | Photonic topological phase transition on the continuum platform through degeneracy-dimension hierarchy.** Adding a bi-anisotropic term in y-z plane transforms the Dirac metamaterial **a** into a photonic topological insulator **b**, where red dashed lines indicate gapless interface states aross the gap. Breaking either inversion symmetry P or time-reversal symmetry T splits Dirac points **a** into Weyl points **g** and **h**, respectively. **c** and **d** indicate two transient phases, where each triple degenerate point (TDP) consists of two longitudinal modes and one circular polarized transverse mode. After shifting the two longitudinal modes (ShiftL) away from each other, **c** and **d** change to be **g** and **h**, respectively. The TDPs in **e** and **f** are neutral, with

imposing only $C_2$ rotation symmetry along z, they turn to be nodal lines **i**. Hollow square, triangle and ellipsoid represent four-fold, triple and double degeneracy, respectively. Colorful (black) symbols indicate chiral (neutral) gapless excitations. The gray/black-dashed, red/blue and cyan/magnental lines indicate mutually orthogonal: electric/magnetic-longitudinal (E-L/H-L), left/right circular polarized (LCP/RCP) transverse and linear polarized (TE/TM) transverse modes, respectively. Corresponding hypothetical constitutive relation examples for panel **a-i** are given in Table. S1.

## ACKNOWLEDGMENTS

This work was financially supported by the European Research Council Consolidator Grant (Topological), Horizon 2020 Action Project grant 734578 (D-SPA), Leverhulme Trust (grant RPG-2012-674) and Research Grants Council Hong Kong (16304717, AoE/P-02/12). S.Z. acknowledges support from the Royal Society and Wolfson Foundation. J. C. acknowledges the financial support National Natural Science Foundation of China (11574162, 11874228) Q.G. acknowledges the financial support of the National Natural Science Foundation of China (grant 11604216).